\documentclass[aps,prd,showpacs,floatfix,citeautoscript,longbibliography,superscriptaddress]{revtex4}
\usepackage{graphicx}
\usepackage{bm,amsmath,amssymb,mathrsfs,dcolumn}
\usepackage[colorlinks=true, linkcolor=blue, citecolor=blue, urlcolor=blue, linktoc=page, bookmarks=false, pdfstartview={FitH}, pdfborder={0 0 0.0 [3 3]}]{hyperref} 
\usepackage[usenames]{color}
\hypersetup{pdfborder=0 0 0,colorlinks=true,citecolor=blue,linkcolor=blue}
\usepackage[final]{pdfpages}
\usepackage{epstopdf}
\usepackage{subfigure}
\usepackage{units}
\usepackage{lineno}
\usepackage{url}
\usepackage{multirow}
\usepackage{graphicx}
\usepackage{booktabs}

\newcommand{\T}{$\sin^22\theta_{13}$}
\newcommand{\U}{$^{238}$U}
\newcommand{\Th}{$^{232}$Th}

\begin{document}

\title{Radiogenic neutron background in reactor neutrino experiments}
\author{Zhiyuan Chen}
\affiliation{Institute of High Energy Physics, Chinese Academy of Sciences, Beijing 100049, China}
\affiliation{University of Chinese Academy of Sciences, Beijing 100049, China}

\author{Xin Zhang}
\affiliation{Institute of High Energy Physics, Chinese Academy of Sciences, Beijing 100049, China}
\affiliation{University of Chinese Academy of Sciences, Beijing 100049, China}

\author{Zeyuan Yu}
\altaffiliation{Corresponding author: yuzy@ihep.ac.cn}
\affiliation{Institute of High Energy Physics, Chinese Academy of Sciences, Beijing 100049, China}

\author{Jun Cao}
\affiliation{Institute of High Energy Physics, Chinese Academy of Sciences, Beijing 100049, China}
\affiliation{University of Chinese Academy of Sciences, Beijing 100049, China}

\author{Changgen Yang}
\affiliation{Institute of High Energy Physics, Chinese Academy of Sciences, Beijing 100049, China}
\affiliation{University of Chinese Academy of Sciences, Beijing 100049, China}

\date{\today }

\begin{abstract}
We report a novel correlated background in the antineutrino detection using the inverse beta decay reaction.
Spontaneous fissions and $(\alpha,n)$ reactions in peripheral materials of the antineutrino detector, such as borosilicate glass of photomultipliers, produce fast neutrons and prompt gamma rays.
If the shielding from the material to the detector target were not thick enough, neutrons and gammas could enter the target volume and mimic antineutrino signals.
This paper revisits the yields and energy spectra of neutrons produced in B$(\alpha,n)$N and F$(\alpha,n)$Na reactions.
A Geant4 based simulation has been carried out using a simplified detector geometry for the present generation reactor neutrino experiments.
The background rates in these experiments are estimated.
If this background was not taken into account, the value of the neutrino mixing angle \T~would be underestimated.
We recommend that Daya Bay, RENO, Double Chooz, and JUNO, carefully examine the masses and radiopurity levels of detector materials that are close to the target and rich in boron and fluorine.

\end{abstract}


\maketitle

\section{Introduction}
Neutrino oscillation is the only phenomenon beyond the Standard Model of particle physics with solid evidence at present.
Reactor antineutrinos have been playing an important role in studies of the neutrino oscillation, including the measurement of the neutrino mixing angle $\sin^2\theta_{12}$ and the squared mass split $\Delta m^2_{21}$ at KamLAND~\cite{Gando:2013nba}, the observation of the neutrino oscillation driven by the mixing angle $\theta_{13}$ at Daya Bay~\cite{2012DYB}, RENO~\cite{2012RENO}, and Double Chooz~\cite{2012Chooz}.
Determination of the neutrino mass ordering at JUNO will also use reactor antineutrinos~\cite{An:2015jdp}.
In the following sections, reactor antineutrinos will be called reactor neutrinos for brevity.

The precise measurement of $\theta_{13}$ is a key to understand the flavor and mass structure of nature.
It is also a crucial input for future experiments to determine the neutrino mass hierarchy and search for neutrino CP violation.
In reactor neutrino experiments, $\theta_{13}$ is measured primarily by the deficit of neutrino rates and spectral distortion observed in near and far detectors.
Detection of reactor neutrinos uses the inverse $\beta$-decay (IBD) reaction in liquid scintillator, $\bar{\nu}_e + p \rightarrow e^+ + n$.
The coincidence of the prompt scintillation from the $e^+$ and the delayed signal from the neutron capture on gadolinium~(nGd) or hydrogen~(nH) reduces the radioactive background significantly.
The amount of accidental background, which is formed by two uncorrelated interactions accidentally satisfying the antineutrino selection criteria, can be estimated with negligible uncertainty.
Thus, a crucial task in these experiments is to identify correlated background from physical processes that produce a pair of correlated interactions that potentially mimic inverse beta decay.

There have been many studies on the correlated backgrounds in reactor neutrino experiments.
The backgrounds consist of cosmogenic $^9$Li and $^8$He isotopes, cosmogenic fast neutrons, and intrinsic $^{13}$C($\alpha$, n)$^{16}$O reactions in liquid scintillator.
Geo-neutrinos, which are electron antineutrinos with energies up to about 3.2~MeV and produced in the \U~and \Th~decay chains in the Earth, can be safely ignored in experiments measuring the mixing angle \T.
The additional background could be present in a particular experiment, such as the background from Am-C neutron calibration sources of Daya Bay~\cite{Gu:2015inc} and the $^{252}$Cf contamination background of RENO~\cite{Kim:2016yvm}.
A common feature of these correlated backgrounds is the appearance of neutrons.
In addition to the above mentioned physics processes, neutrons could be produced in peripheral materials via spontaneous fissions, ($\alpha$, n) reactions, and photon nuclear processes.
These neutrons are defined as radiogenic neutrons in peripheral materials hereafter, and will be called radiogenic neutrons for brevity.

Although radiogenic neutrons have been studied intensively in solar neutrino experiments and in dark matter experiments, their contributions to the detection of antineutrinos have not been discussed yet.
In solar neutrino experiments, radiogenic neutrons could be captured on peripheral detector components, releasing high-energy gamma rays.
Once a gamma ray deposits enough energy in the target, it could mimic the scattered electron signal at Borexino~\cite{Agostini:2017cav}, and mimic the neutral current signal at SNO~\cite{Aharmim:2011vm}.
In experiments searching for weak interactions massive particles~(WIMPs), simulating radiogenic neutrons in every piece of shielding and detector material is of greater importance~\cite{Westerdale:2020iqy, Westerdale:2017kml, Cooley:2017ocy, Mendoza:2019vgf,Vlaskin:2015hhf}, because a nuclear recoil from a neutron cannot be distinguished from a nuclear recoil from a WIMP.

We have recently discovered that radiogenic neutrons produced in peripheral materials could form correlated backgrounds in reactor neutrino experiments.
In this work, we have revisited the $(\alpha, n)$ yields on two crucial elements, boron and fluorine.
Compared to previous calculations in literature, detailed information is provided for reactions in which the final nucleus is populated to an excited state.
Then, a Geant4 based simulation was carried out using a simplified detector geometry for the present generation reactor neutrino experiments.
We found that the background cannot be neglected in the neutrino dataset detected via neutron capture on hydrogen of Daya Bay.
The impact on the determination of $\sin^{2}2\theta_{13}$ is discussed.

The paper is constructed as follows: in Sec.~\ref{sec2}, the radiogenic neutron yields and energy spectra are described.
The background simulation in a detector like Daya Bay is presented in Sec.~\ref{sec3}.
Finally, Sec.~\ref{sec5} is devoted to the conclusion.

\section{Yields and energy spectra of radiogenic neutrons}
\label{sec2}

\begin{figure}[htb]
\centering
\includegraphics[width=12cm]{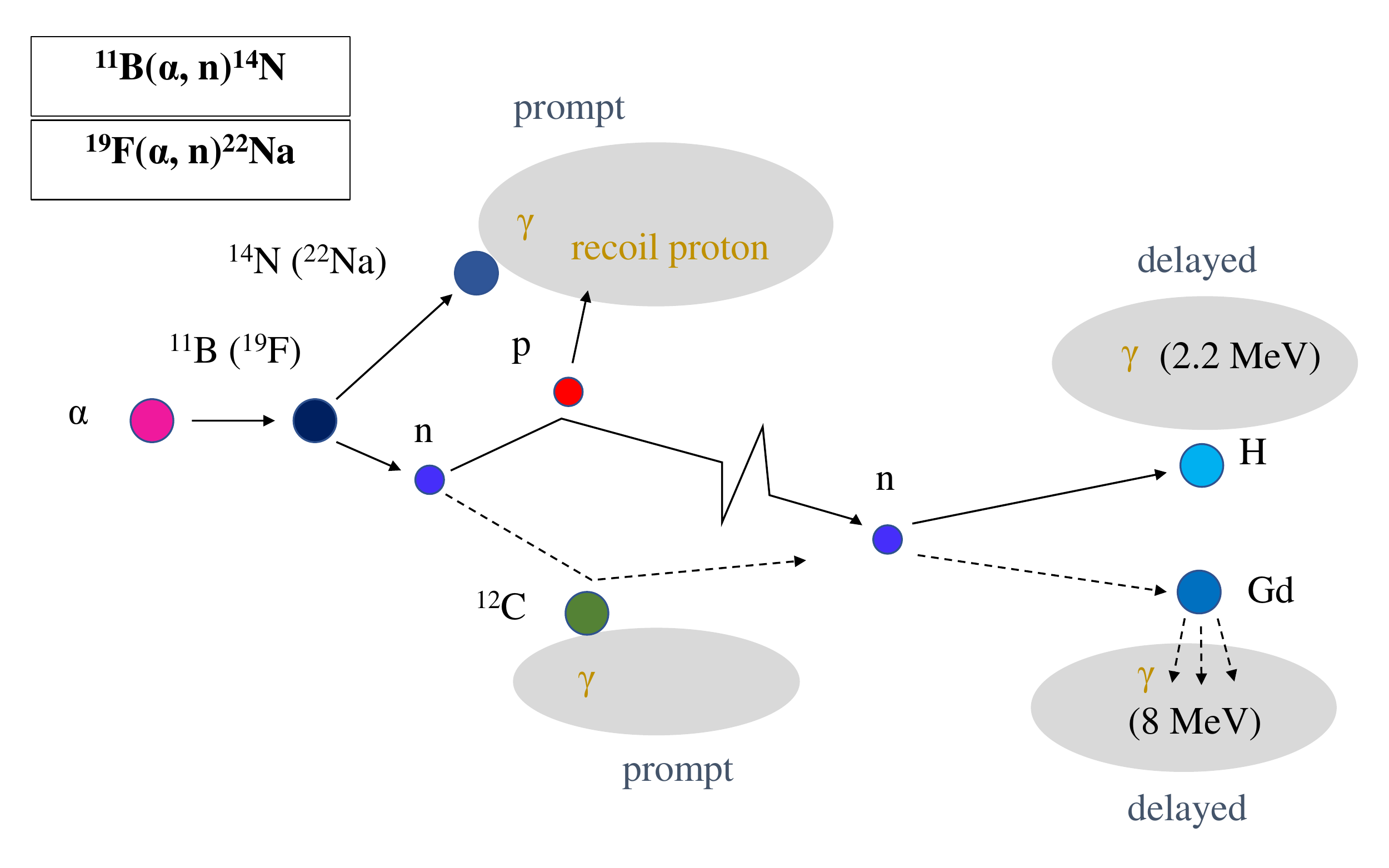}
\caption{Scheme of the correlated background due to ($\alpha,n$) reactions in peripheral materials.}
\label{fig:AlphaNBkg}
\end{figure}

\begin{figure}[htb]
\centering
\includegraphics[width=12cm]{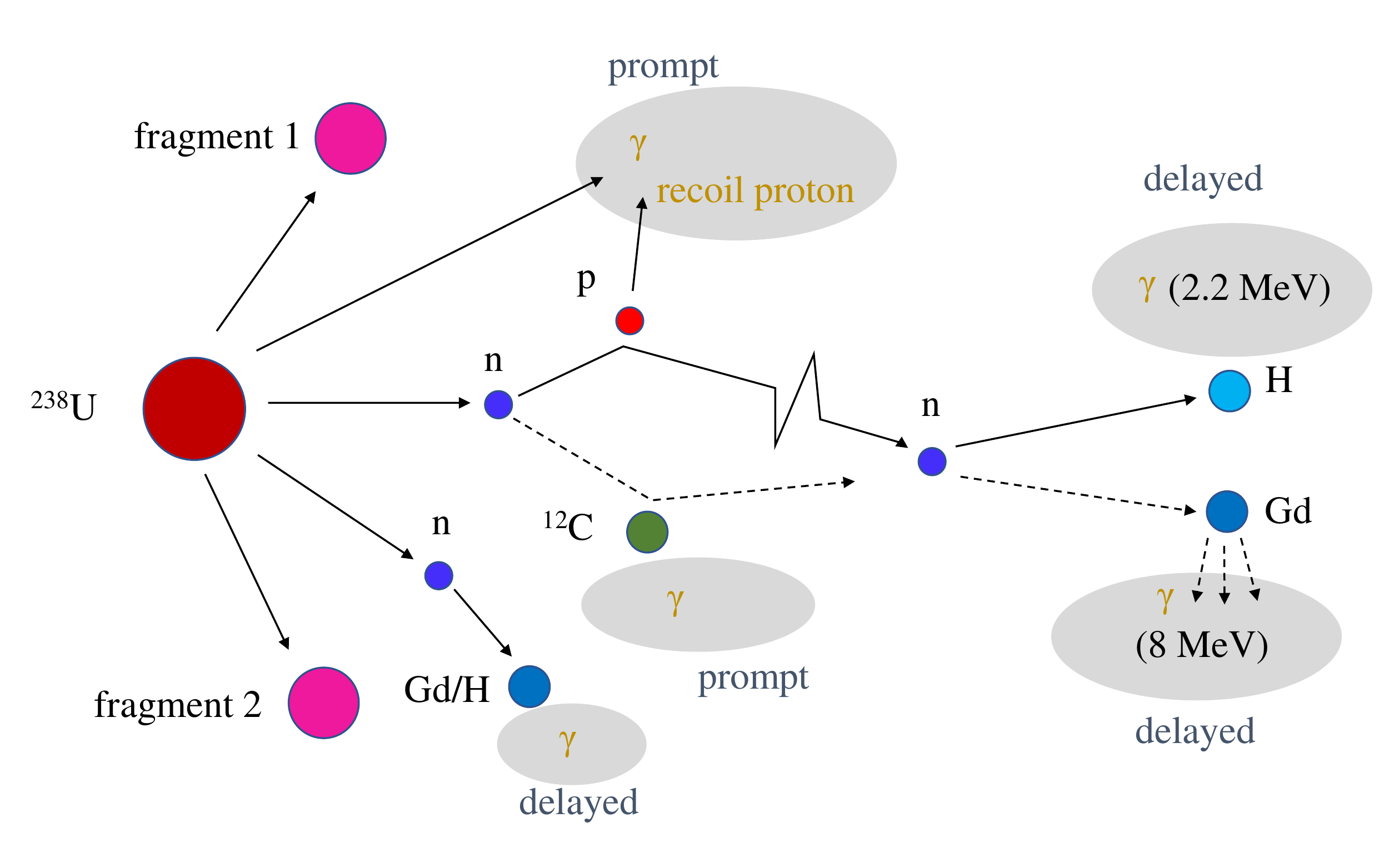}
\caption{Scheme of the correlated background due to \U~spontaneous fissions in peripheral materials.}
\label{fig:FissionBkg}
\end{figure}

Most radiogenic neutrons are produced in ($\alpha$, n) reactions and in spontaneous fissions in low background experiments.
If the shielding between peripheral materials and the detector target were not thick enough, neutrons and $\gamma$ rays could deposit energy in the target volume.
For ($\alpha$,~n) reactions, protons scattered by the neutron or de-excitation gamma rays from excited states of $^{12}$C could form a prompt signal, while the delayed signal comes from the eventual neutron capture.
In the case of spontaneous fissions, several prompt gamma rays and fast neutrons are released simultaneously.
In addition to the cases mentioned above, the correlation may even happen between multiple neutron captures.
Illustrations of these two types of correlated background are shown in Fig.~\ref{fig:FissionBkg} and Fig.~\ref{fig:AlphaNBkg}, respectively.

Among the naturally fissile isotopes, \U~dominates the neutron yields due to its relatively large fission ratio compared to other isotopes.
For ($\alpha$, n) reactions, among the commonly used detector materials, the primary neutron production channels are $\alpha+$B reactions in borosilicate glass and $\alpha+$F reactions in any material rich in fluorine, such as Viton and polytetrafluoroethylene~(PTFE).
In this section, we will present the neutron yields and energy spectra used in our study.

\subsection{Neutrons from spontaneous fissions}

\U, a natural radioactive isotope of great importance in low background experiments, occasionally decays by spontaneous fission with a probability of 5.5$\times 10^{-7}$.
In each spontaneous fission, an energy of about 200~MeV is released.
Most of the released energy is carried by kinetic energies of fission fragments.
The fission process often produces free neutrons and gamma rays.
For fissions in peripheral materials, fission fragments are stopped in these materials, but neutrons and gamma rays could enter the target volume, such as liquid scintillator, and mimic neutrino signals.
The prompt signal could be the combination of prompt gamma rays, neutron recoils on proton, or a gamma ray from early neutron capture.
The delayed signal is the gamma from the eventual neutron capture.
Thus, the angular correlation of fission neutrons and gammas could affect the background rate.
In summary, the following key parameters should be considered:
\begin{itemize}
 \item{Neutron multiplicity per fission.}
 \item{Neutron energy spectrum.}
 \item{Angular correlation of fission neutrons.}
 \item{Gamma multiplicity per fission.}
 \item{Gamma energy spectrum.}
 \item{Correlation between gamma rays and neutrons. }
\end{itemize}

\begin{figure}[htb]
\centering
\includegraphics[width=12cm]{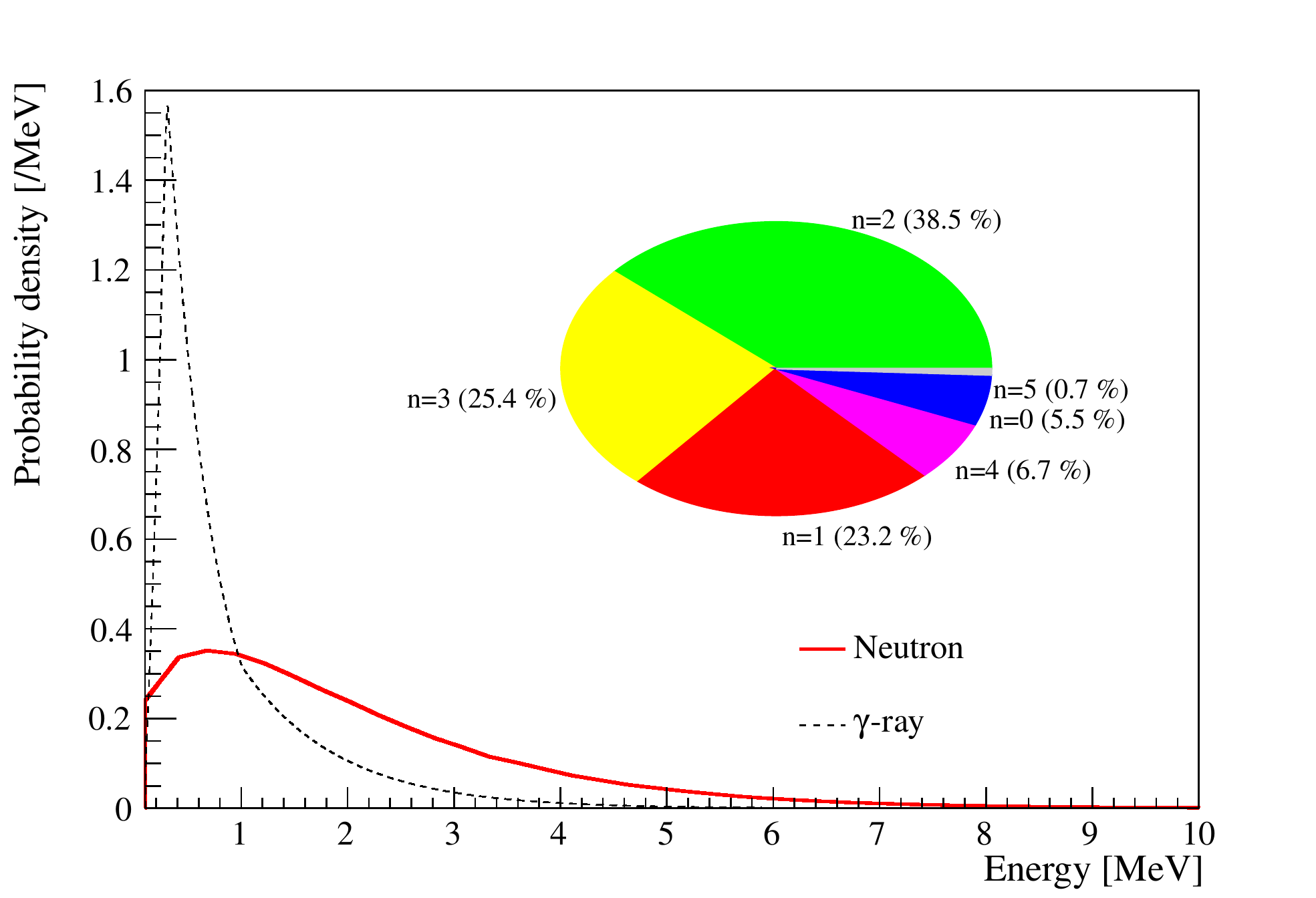}
\caption{Kinetic energy spectrum of neutrons from $^{238}$U spontaneous fissions~(red solid line), the energy spectrum of prompt gamma rays~(dashed blue line). The inserted chart shows the distribution of neutron multiplicity. The numbers are taken from Ref.~\cite{LLNLFission}.}
\label{U238SF}
\end{figure}

The complicated output is well handled by a mature fission generator FREYA developed by LLNL~\cite{FREYA}.
We summarize the elemental distributions obtained from the generator in Fig.~\ref{U238SF}.
On average, about two neutrons are produced per spontaneous fission of \U, and the mean kinetic energy is about 1.8~MeV.
Prompt gamma rays carry a total energy of about 6.9~MeV, and the average number of gamma is 6.3.
The generator outputs are used in the following simulation in Sec.~\ref{sec3}.

\subsection{Neutrons from $(\alpha,n)$ reactions }

\begin{figure}[h]
\centering
\includegraphics[width=12cm]{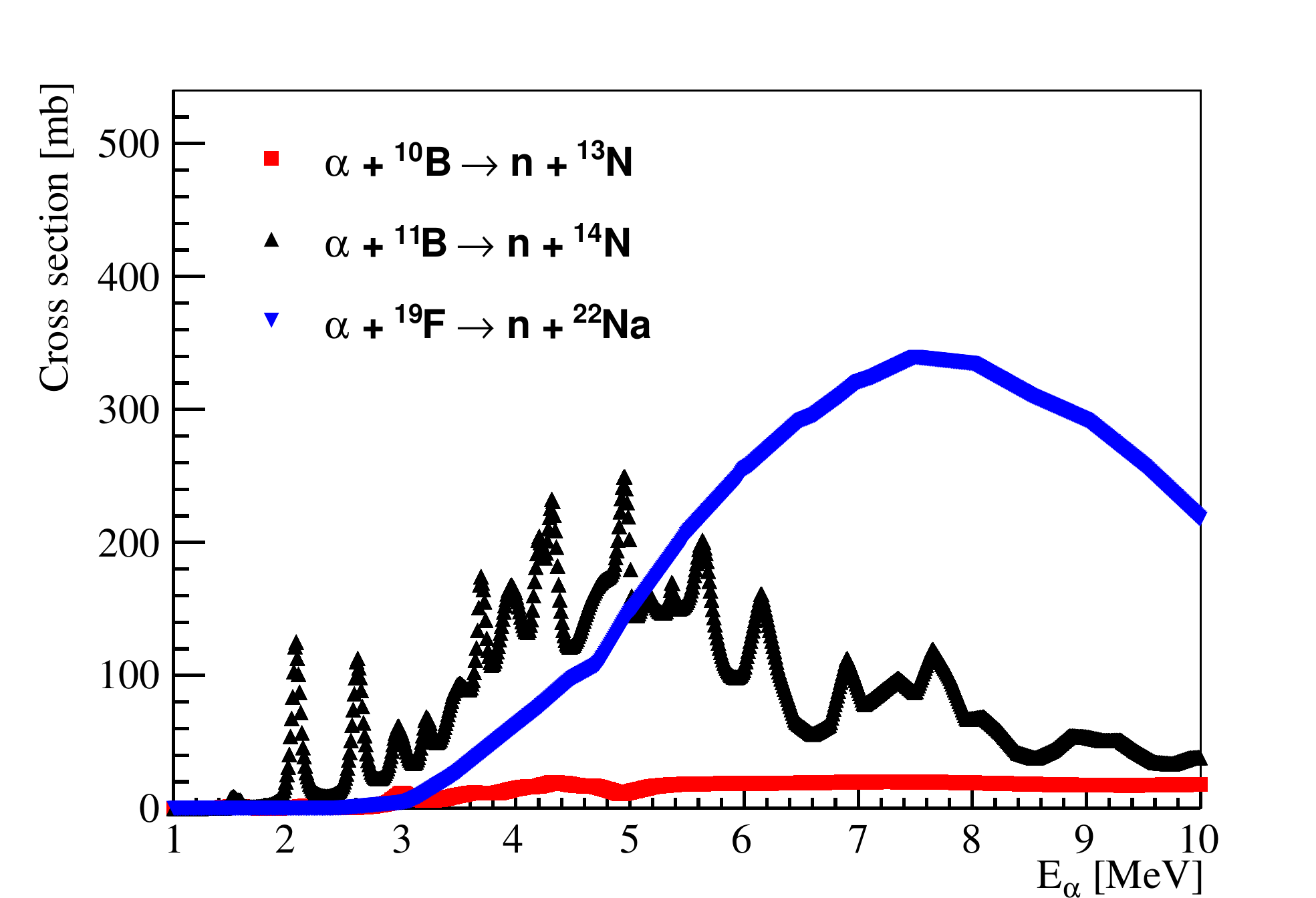}
\caption{Cross sections of $^{10}$B($\alpha$,~n)$^{13}$N, $^{11}$B($\alpha$,~n)$^{14}$N, and $^{19}$F($\alpha$,~n)$^{22}$Na. Values are taken from JENDL-AN/2005~\cite{JENDL}.}
\label{fig:alphaBFTotal}
\end{figure}

In detector materials, most of alpha particles is from the decay chains of \U~and \Th.
There are 8 and 6 alpha particles released when they decay to $^{206}$Pb and $^{208}$Pb, respectively.
The kinetic energies of these $\alpha$s are less than 10~MeV.
The two decay chains are assumed to be in secular equilibrium to simplify the calculation.
However, it is worthy to mention that such equilibrium is rarely achieved in reality.
When an $\alpha$ is generated in a material, its energy is lost quickly via ionization.
However, there is a small probability of inelastically scattering between an alpha particle and a nucleus.
In this case, secondary particles could be generated, such as neutron, proton, gamma ray.
As a crucial neutron source in solar neutrino experiments and in dark matter experiments, the neutron yields in various $(\alpha,n)$ reactions have been calculated~\cite{Westerdale:2020iqy, Westerdale:2017kml, Cooley:2017ocy, Mendoza:2019vgf,Vlaskin:2015hhf}.
Known from these studies, three reactions dominate the neutron yields:
\begin{itemize}
 \item{$\alpha+^{10}$B $\rightarrow$ n $+^{13}$N, $Q$=1.06~MeV;}
 \item{$\alpha+^{11}$B $\rightarrow$ n $+^{14}$N, $Q$=0.16~MeV;}
 \item{$\alpha+^{19}$F $\rightarrow$ n $+^{22}$Na, $Q$=-1.95~MeV.}
\end{itemize}
In these calculations, two databases are frequently used to get the reaction cross sections, the evaluated database JENDL-AN/2005~\cite{JENDL}, and the
database TENDL/2019~\cite{Koning:2019qbo} which is based on both default and adjusted TALYS calculations and data from other sources.
The cross sections of the three reactions taken from JENDL-AN/2005 are drawn in Fig.~\ref{fig:alphaBFTotal}.
In reactor neutrino experiments, boron is rich in borosilicate glass of photomultipliers tubes~(PMTs), while fluorine is rich in electronics boards, O-rings, painting materials, PTFE, etc.

The final nucleus could be populated to an excited state in ($\alpha$,~n) reactions.
Thus, the prompt signal could be formed by protons scattered by the neutron or de-excitation gamma rays, while the delayed signal comes from the eventual neutron capture.
In the following, ($\alpha$,~n$_i$) is used to denote the reaction in which the final nucleus is populated to the $i_{\rm th}$ excited state.
Although several calculations of the total neutron yields and energy spectra have been made in literature~\cite{Westerdale:2020iqy, Westerdale:2017kml, Cooley:2017ocy, Mendoza:2019vgf,Vlaskin:2015hhf}, we do not find the output of ($\alpha$,~n$_i$) reactions.

This triggers us to perform the calculation of neutron yield $Y^{\alpha}_i$ for a given $\alpha$ kinetic energy $T_{\alpha}$ by integrating over the entire track length:
\begin{equation}
\begin{aligned}
Y^{\alpha}_i(T_n) = \frac{N_A\cdot C \cdot \rho}{A} \int^{T_{\alpha}}_{0} \frac{\sigma_i(T'_{\alpha},T_n)}{S(T'_{\alpha})}dT'_{\alpha},
\end{aligned}
\label{yield}
\end{equation}
where $i$ denotes the $i_{\rm th}$ excited state of the final nucleus, $T_n$ is the kinetic energy of neutron, $N_A$ is the Avogadro's number, $\rho$ is the density of the material, $C$ and $A$ are the mass fraction and the mass number of the initial nucleus.
$\sigma_i$ is the double differential cross section of this reaction.
$S$ is the mass stopping power calculated using the SRIM code~\cite{ZIEGLER20101818}.
The neutron angular distribution is assumed to be isotropic in the center of mass system.
Thus, the required inputs consist of
\begin{itemize}
 \item{Chemical components of the material,}
 \item{Mass density of the material,}
 \item{Cross section of ($\alpha,n_i$) reactions.}
\end{itemize}

Reactions between alpha particles and the other elements are not discussed in this paper for several reasons.
The first group of elements has high $(\alpha,n)$ reaction thresholds which are above the energies of alpha particles from natural radioactive decays, such as $^{12}$C, $^{16}$O, $^{28}$Si.
The second group has small natural abundance, such as $^{13}$C, $^{17}$O, $^{18}$O, $^{29}$Si and $^{30}$Si.
The last group, such as iron and copper, has low neutron energies that are difficult to pass the IBD analysis threshold 0.7~MeV.

\subsubsection{ B($\alpha$,~n)N reactions}

In borosilicate glass, the mass fraction of boron oxide varies from 8\% to 20\%.
Following Ref.~\cite{Westerdale:2017kml}, a typical value of 18\% is suggested for Hamamatsu customer's reference.
It will be used in the following calculations.
The other 82\% is simplified to be SiO$_2$.
The mass density of borosilicate glass is set to 2.23~g/cm$^3$.
Using these inputs, the stopping power of alpha particles is calculated.
The range of an $\alpha$ particle with a kinetic energy of 5~MeV is 22.3~$\mu$m in the glass.

\begin{figure}[h]
\centering
\includegraphics[width=12cm]{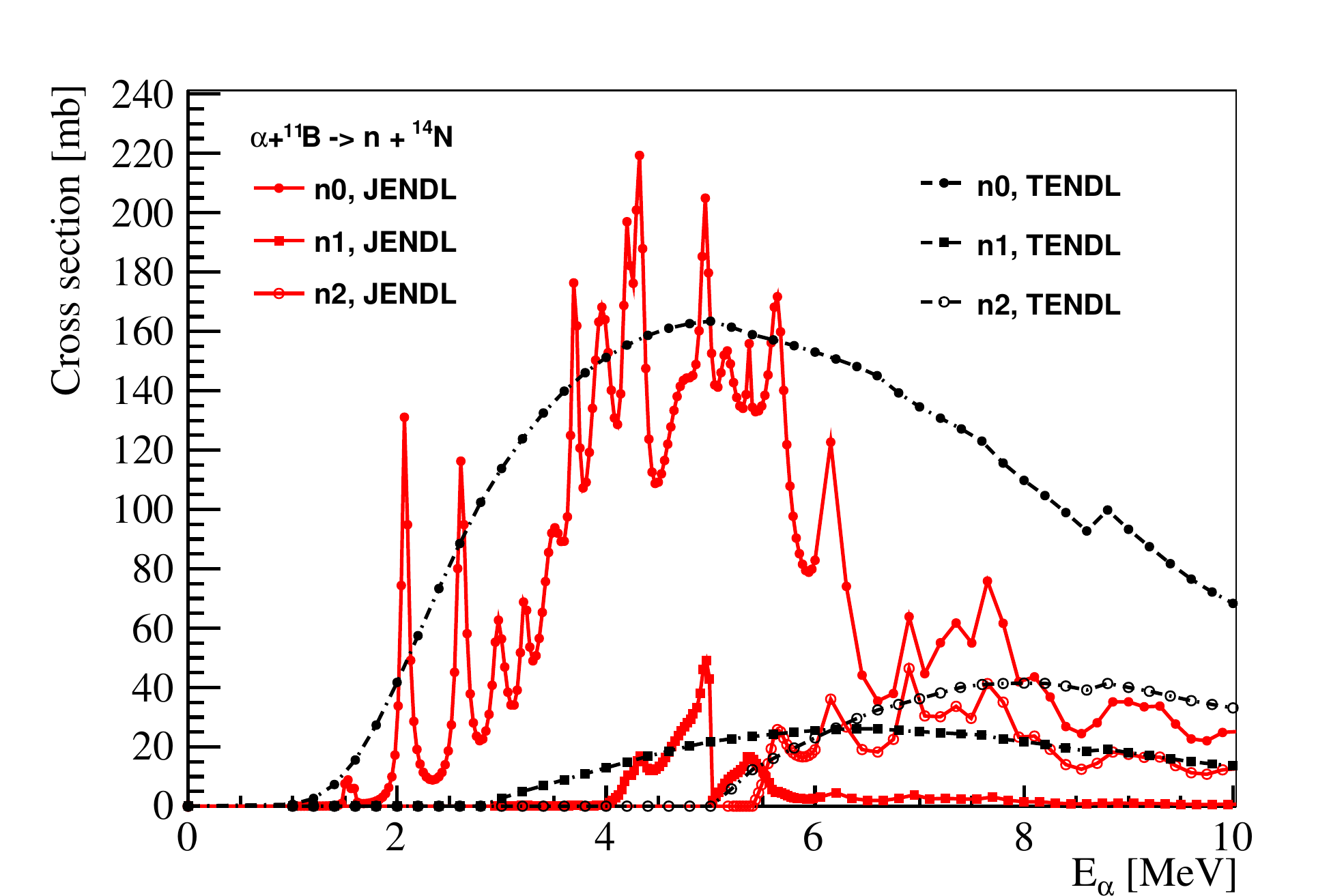}
\caption{Cross sections of $^{11}$B($\alpha$,~n$_{i}$)$^{14}$N reactions. Data taken from JENDL-AN/2005 and TENDL/2019 are drawn for comparison.}
\label{fig:CXcompareB11}
\end{figure}

The cross sections of $^{11}$B($\alpha$,~n$_{i}$)$^{14}$N reactions are shown in Fig.~\ref{fig:CXcompareB11}.
The values taken from TENDL/2019 are larger than those from JENDL-AN/2005.
In addition, many resonances can be found in the latter one.
Since we are not able to justify which database is more accurate, the calculation is performed based on cross sections from both databases.
The difference between them could be treated as systematics.

\begin{figure}[h]
\centering
\includegraphics[width=12cm]{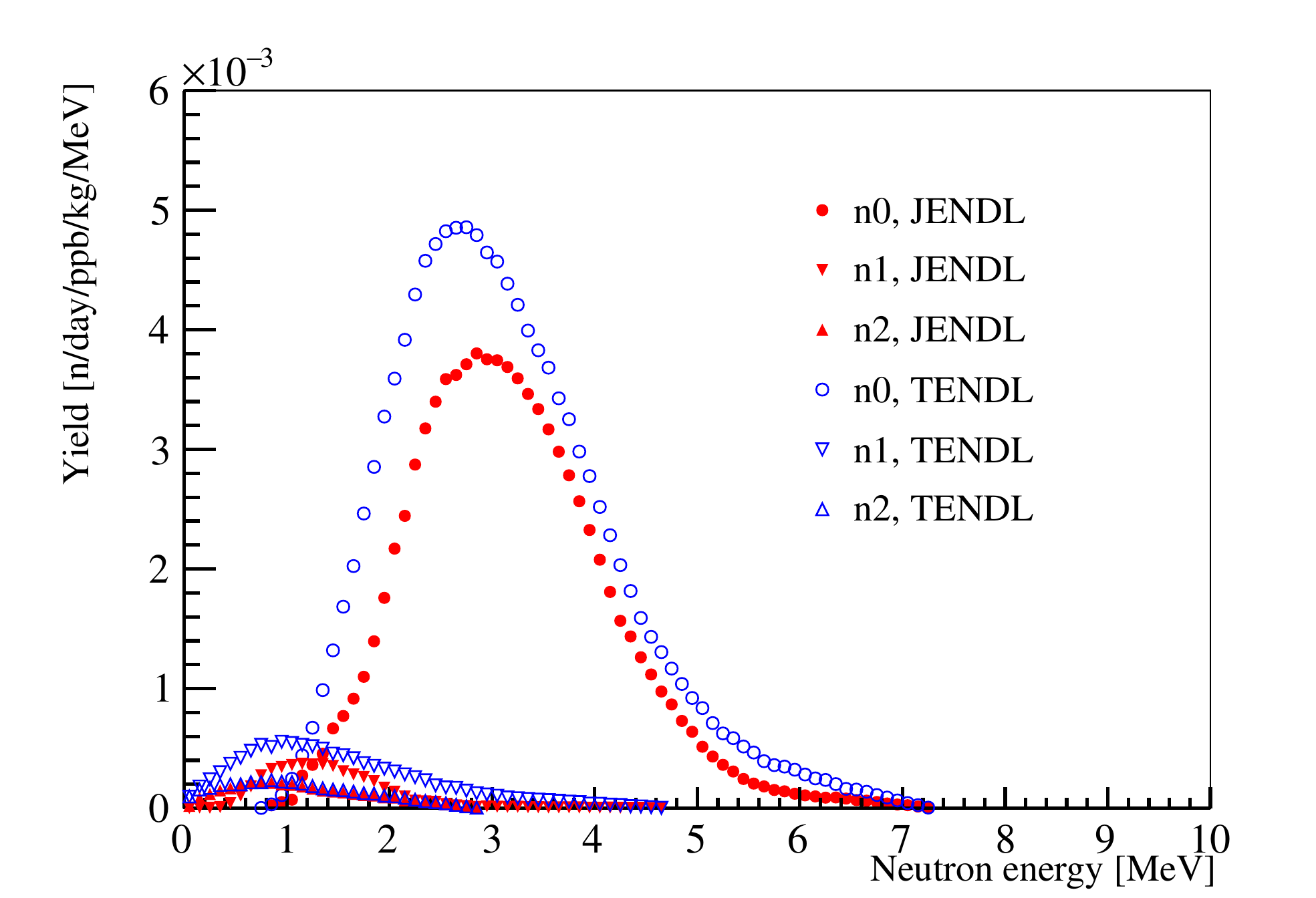}
\caption{Yields and energy spectra of neutrons from $^{11}$B($\alpha$,~n$_{i}$)$^{14}$N reactions in borosilicate glass. The $\alpha$ particles are from the \U~decay chain in secular equilibrium.}
\label{fig:neutronB11}
\end{figure}

In $^{11}$B($\alpha$,~n$_{i}$)$^{14}$N reactions, the final $^{14}$N nucleus could be populated to the first two excited states.
Taking the \U~decay chain in equilibrium as the alpha source, the calculated neutron yields and spectra of these reactions are shown in Fig.~\ref{fig:neutronB11}.
The de-excitation information of $^{14}$N is taken from Ref.~\cite{Ajzenberg-Selove:1991rsl}.
The distribution of neutron energies is peaked at 3~MeV for the ($\alpha$,~n$_{0}$) reaction.
The total neutron yield is 9.6$\times 10^{-3}$ per day per kg per ppb \U~using cross sections from JENDL-AN/2005, and 13.6$\times 10^{-3}$ per day per kg per ppb \U~using TENDL/2019.
Taking the \Th~decay chain in equilibrium as the alpha source, the neutron yields are 3.8$\times 10^{-3}$ per day per kg per ppb \Th~(JENDL-AN/2005) and 5.8$\times 10^{-3}$ per day per kg per ppb \Th~(TENDL/2019).
Deexcitation of $^{14}$N$^{*}$ is via emission of prompt gamma ray(s).
Table~\ref{table:excitedN14} summarizes the probability of each explicit channel and its decay schema.
The higher mean energy of $\alpha$ particles of \Th~chain results in the larger probabilities of reaching excited states of $^{14}$N.

\renewcommand{\arraystretch}{1.2}
\begin{table}[hbt]
\caption{Probabilities of the $^{11}$B($\alpha$,~n$_{i}$)$^{14}$N reactions, calculated using cross sections in JENDL-AN/2005. }
\centering
\begin{ruledtabular}
\begin{tabular}{cccc}
\multirow{2}{*}{$^{14}$N level $i$ } & \multicolumn{2}{c}{Branching ratio (\%)} & \multirow{2}{*}{$\gamma$-rays energy [MeV]} \\ \cline{2-3}
   &$^{238}$U& $^{232}$Th & \\ \hline
0 & 91.4 & 88.4 & 0   \\
1 & 5.1  & 5.5  & 2.2 \\
2 & 3.5  & 6.1  & 2.2 + 1.6 \\
\end{tabular}
\end{ruledtabular}
\label{table:excitedN14}
\end{table}

\begin{figure}[h]
\centering
\includegraphics[width=12cm]{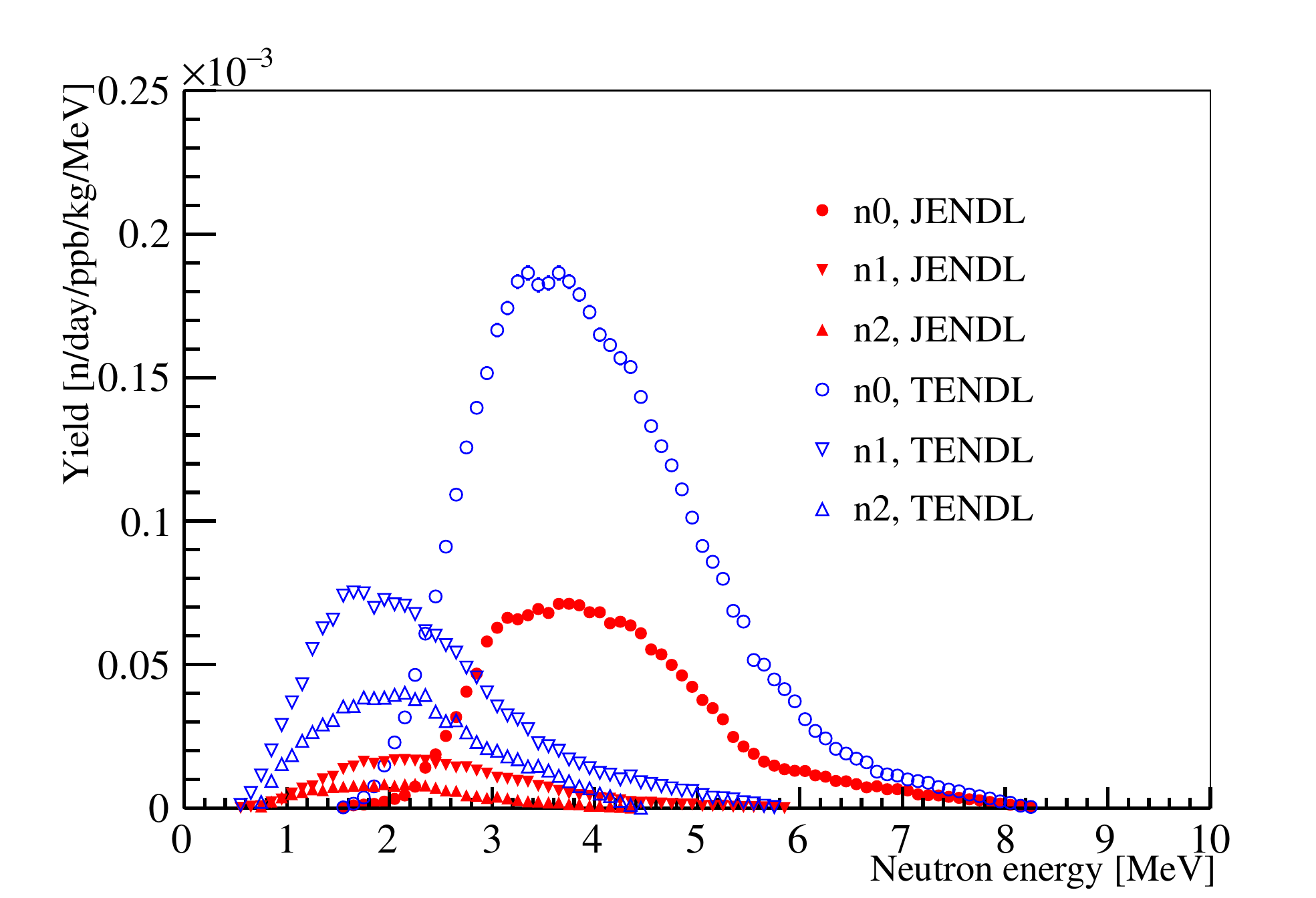}
\caption{Yields and energy spectra of neutrons from $\alpha$ + $^{10}$B $\rightarrow$ $^{13}$N + n ($^{12}$C + n + $p$) reactions in borosilicate glass. The bombarded $\alpha$ particles are from the \U~decay chain in secular equilibrium. }
\label{B10SpecAlphaN}
\end{figure}

The $^{10}$B($\alpha$,~n$_{i}$)$^{13}$N reaction is more complicated.
Unlike $^{14}$N, a proton with a kinetic energy of several MeV would be emitted when $^{13}$N is at the excited states~\cite{Ajzenberg-Selove:1991rsl}.
The proton stops quickly in the peripheral material and has no contribution to the background.
Thus, only neutrons are considered in the study, and the energy spectra are shown in Fig.~\ref{B10SpecAlphaN}.
The total neutron yield is 2.0$\times 10^{-4}$ per day per kg per ppb \U~using cross sections from JENDL-AN/2005, and 7.6$\times 10^{-4}$ per day per kg per ppb \U~using TENDL/2019.
Taking the \Th~decay chain in equilibrium as the alpha source, the neutron yields are 1.0$\times 10^{-4}$ per day per kg per ppb \Th~(JENDL-AN/2005) and 3.2$\times 10^{-4}$ per day per kg per ppb \Th~(TENDL/2019).
Table~\ref{table:excitedN13} summarizes neutron yields of $^{10}$B($\alpha$,~n$_{i}$)$^{13}$N reactions and their decay schema.

\renewcommand{\arraystretch}{1.2}
\begin{table}[hbt]
\caption{Probabilities of $^{10}$B($\alpha$,~n$_{i}$)$^{13}$N reactions, calculated using cross sections in JENDL-AN/2005. }
\centering
\begin{ruledtabular}
\begin{tabular}{ccc}
\multirow{2}{*}{$^{13}$N level $i$} & \multicolumn{2}{c}{Branching ratio (\%)}   \\ \cline{2-3}
   &$^{238}$U& $^{232}$Th  \\ \hline
0 & 67.0 & 59.3   \\
1 & 22.0  & 22.7   \\
2 & 11.0  & 18.0   \\
\end{tabular}
\end{ruledtabular}
\label{table:excitedN13}
\end{table}

\subsubsection{ $^{19}$F($\alpha$,~n)$^{22}$Na reaction}

\begin{figure}[h]
\centering
\includegraphics[width=12cm]{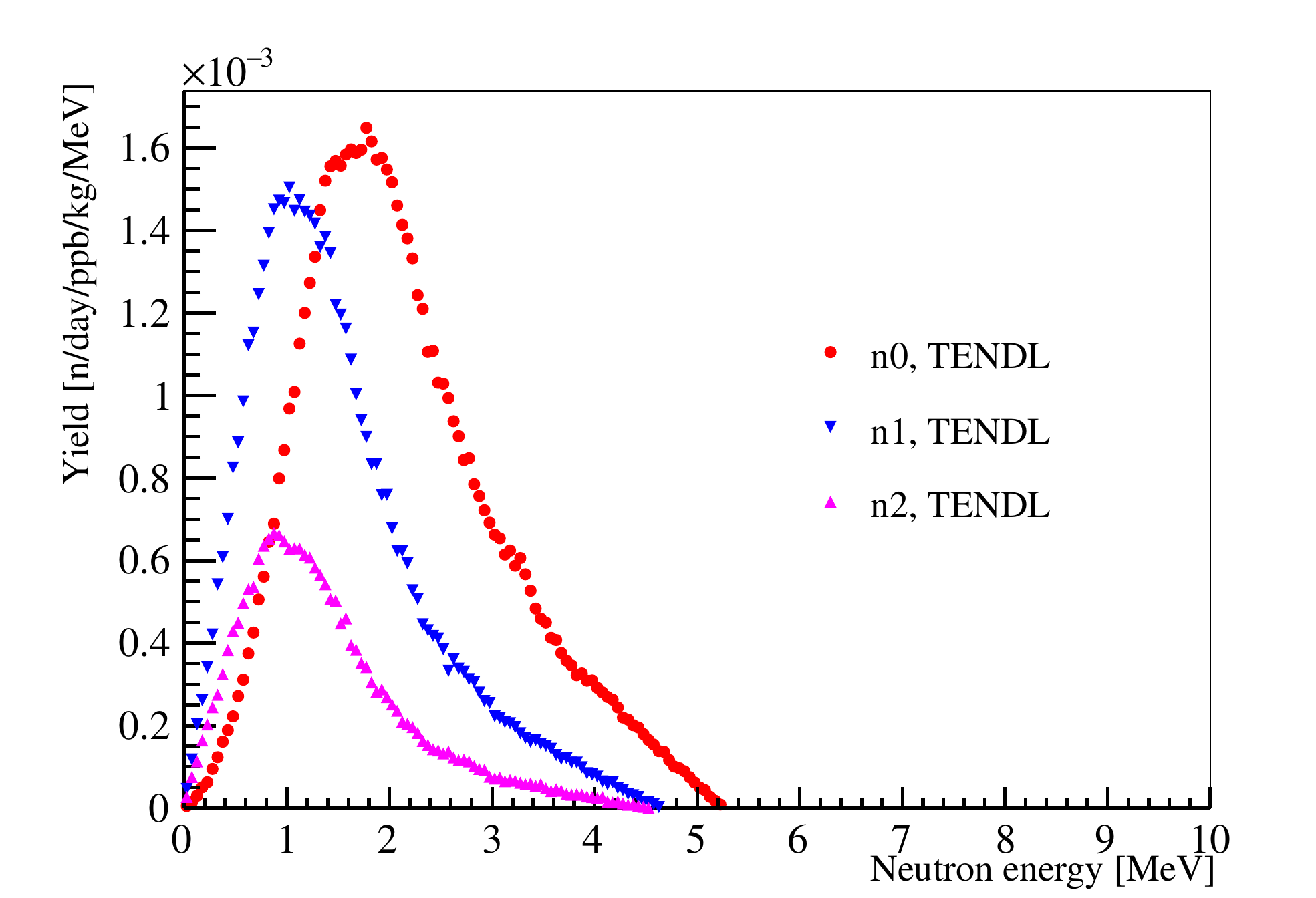}
\caption{Yields and energy spectra of neutrons from $^{19}$F$(\alpha,n_{0,1,2})^{22}$Na reactions in PTFE.}
\label{F19SpecAlphaN}
\end{figure}

There are a lot of materials rich in fluorine used in neutrino detectors.
PTFE~(C$_2$F$_4$, mass density 1.7~g/cm$^3$) is taken as an example in this study.
For this reaction, the total and explicit cross sections from JENDL-AN/2005 and from TENDL/2019 agree to 5\%.
Thus, only cross sections from TENDL/2019 are used.
There are fifteen states of $^{22}$Na that can be populated in this reaction as listed in Table~\ref{excitedNa22}.
For $\alpha$ particles from $^{238}$U and $^{232}$Th decay chains, the probabilities of reaches these $^{22}$Na states are also provided.
The de-excitation information of $^{22}$Na is taken from Ref.~\cite{Basunia:2015qqx}.
For illustration, spectra of neutrons from $^{19}$F$(\alpha,n_{0,1,2})^{22}$Na reactions are shown in Fig.~\ref{F19SpecAlphaN}.
The total neutron yield is 7.2$\times 10^{-3}$ per day per kg per ppb \U, 5.3$\times 10^{-3}$ per day per kg per ppb \Th~using cross sections from TENDL/2019.

\begin{table}[h]
\caption{Probabilities of $^{19}$F($\alpha$,~n$_{i}$)$^{22}$Na reactions, calculated using cross sections in TENDL/2019.}
\centering
\begin{ruledtabular}
\begin{tabular}{cccc}
\multirow{2}{*}{$^{22}$Na level $i$} & \multicolumn{2}{c}{Probability (\%)} & \multirow{2}{*}{$\gamma$-rays energy [MeV]} \\ \cline{2-3}
   &$^{238}$U& $^{232}$Th & \\ \hline
 0 & 26.9 & 22.9 & 0   \\
 1 & 20.1 & 15.9 & 0.58 \\
 2 &  8.3 &  6.5 & 0.58 + 0.074 \\
 3 & 12.0 & 12.7 & 0.89 \\
 4 &  5.3 &  6.5 & 1.53 \\
 5 &  4.6 &  5.3 & 1.28 + 0.58 + 0.074 \\
 6 &  6.2 &  7.4 & 1.37 + 0.58 \\
 7 &  5.8 &  7.3 & 1.4 + 0.58 \\
 8 &  3.7 &  4.7 & 1.55 + 0.58 + 0.074 \\
 9 &  2.9 &  3.9 & 2.57 \\
10 &  1.7 &  2.3 & 1.01 + 1.37 + 0.58 \\
11 &  1.3 &  1.8 & 1.1 + 1.37 + 0.58 \\
12 &  0.9 &  1.6 & 1.59 + 1.37 + 0.58 \\
13 &  0.2 &  0.6 & 2.81 + 0.89 \\
14 &  0.1 &  0.4 & 3.28 + 0.58 + 0.074 \\
\end{tabular}
\end{ruledtabular}
\label{excitedNa22}
\end{table}

The above total neutron yields in borosilicate glass and PTFE are also compared with calculations in literature~\cite{Westerdale:2020iqy, Westerdale:2017kml, Cooley:2017ocy, Mendoza:2019vgf,Vlaskin:2015hhf}.
Good consistency is found.

\section{Radiogenic neutron background in reactor neutrino experiments}
\label{sec3}

In the present generation of reactor neutrino experiments, Daya Bay, RENO, and Double Chooz, the neutrino mixing angle $\theta_{13}$ is measured primarily by comparing the neutrino reaction rates in the near and far detectors.
If there was a residual background with equal rates in near and far detectors, the size of the neutrino disappearance, as well as the value of $\theta_{13}$, would be underestimated.
We take the Daya Bay experiment as an illustration.
All the information used in this section is from published materials in the literature.

\subsection{Simulation}

\begin{figure}[h]
\centering
\includegraphics[width=12cm]{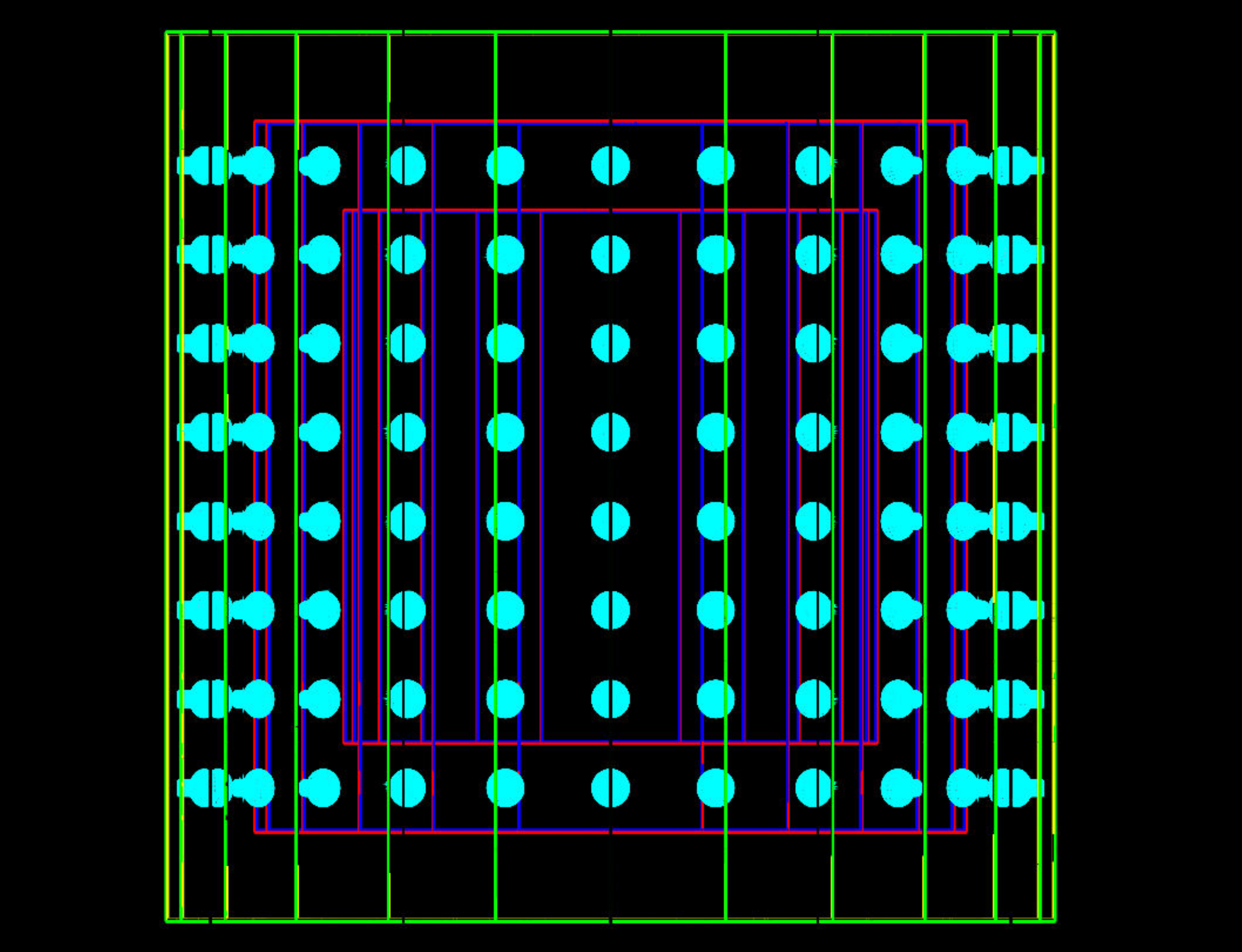}
\caption{Detector geometry of Daya Bay in our simulation using Geant 4.10.2. Green lines are the stainless steel vessel. Red lines are the two acrylic vessels. The light blue tubes are the 192 PMTs installed on the stainless steel vessel.}
\label{fig:detector}
\end{figure}

Following the detector description in Ref.~\cite{DayaBay:2015kir}, we have built a simplified Daya Bay detector in Geant 4.10.2 as shown in Fig.~\ref{fig:detector}.
The detector consists of a cylindrical vessel made of stainless steel and two acrylic vessels.
The heights and diameters of the three vessels are (5~m, 5~m), (4~m, 4~m), (3~m, 3~m), respectively.
The three volumes are filled with mineral oil, liquid scintillator, and Gd-loaded liquid scintillator from the outermost to the innermost.
192 Hamamatsu R5912 8-inch PMTs are located in the inner wall of the stainless steel vessel.
The geometry of the PMT is taken from the datasheet of Hamamatsu~\cite{PMTSheet}.
The distance from the top of the bulb to the LS is only 20~cm.
In the simulation, radiogenic neutrons are generated uniformly in the glass.
Assuming the weight of borosilicate glass is 700~g per PMT, and taking the concentrations of U~(150~ppb) and Th~(350~ppb) from Ref.~\cite{DayaBay:2015kir}, Table~\ref{table:simulation} summarizes the number of simulated reactions.
There are 400 to 600 neutrons produced per day in the glass of the 192 PMTs in one detector.

\begin{table}[htb]
\caption{Number of neutrons and background rates in our simulation in one Daya Bay detector. The (\U$_\alpha,n$) and (\Th$_\alpha,n$) means the bombarded alpha particles are released by \U~and \Th~decay chains, respectively. The JENDL and TENDL means the cross sections of ($\alpha,~n$) reactions are taken from JENDL-AN/2005 and TENDL/2019, respectively.  }
\centering
\begin{ruledtabular}
\begin{tabular}{cccc}
Per day & (\U$_\alpha,n$)  & (\Th$_\alpha,n$)  & $^{238}$U SF \\ \hline
Reactions~(JENDL) &  197 & 183 & 12 \\
Reactions~(TENDL) &  290 & 287 & 12 \\
Bkg. in nH~(JENDL) & 0.037 & 0.034  & 0.06  \\
Bkg. in nH~(TENDL) & 0.054  & 0.052 & 0.06 \\
Bkg. in nGd~(TENDL)& 0.005 & 0.005 & 0.006\\
\end{tabular}
\end{ruledtabular}
\label{table:simulation}
\end{table}

In the simulation, only the particle tracking and energy deposits are included.
Neither the optical and electronics simulation nor the event reconstruction is performed.
To separate the prompt signal, which consists of neutron recoils and prompt gammas, and the delayed signal, i.e., the neutron capture, an event is defined as steps of energy deposits in a time window of 1~$\mu$s.
Variables used in the selection for correlated event pairs are defined as:
\begin{itemize}
\item {Energy: the sum of quenched deposit energies of steps in one event. In each step, the energy deposit is quenched according to the Birks' law~\cite{BirksBook}. The Birks' constant kB is set to 6.5$\times 10^{-3}$~g/cm$^2$/MeV.}
\item {Vertex: average of interaction positions of all steps, weighted by quenched deposit energies.}
\item {Time: average of interaction time of all steps, weighted by quenched deposit energies.}
\end{itemize}

Following the event selection criteria in the Daya Bay publications~\cite{2012DYB,2016nH}, the cuts in this study are listed in Table~\ref{IBDSelection}.
Applying these cuts to the simulation data, the background rates are summarized in Table~\ref{table:simulation}.
In the nH analysis, a background rate of 0.13 and 0.17 per day is found when using ($\alpha,~n$) cross sections from JENDL-AN/2005 and TENDL/2019, respectively.
Although there are only 12 spontaneous fissions per day, their contributions to backgrounds are similar to ($\alpha,~n$) reactions.
This is because the energy carried by gamma rays is higher in spontaneous fissions.
In addition, the correlation may happen between two neutron captures.
Comparing the background rate to the 48 IBD signal per day in one far detector of Daya Bay~\cite{2016nH}, the radiogenic neutron background must be taken into consideration.
In the nGd data set, the background rate is only 0.02 per day, which can be safely neglected.

\begin{table}[hbt]
\caption{IBD selection criteria for the nH~\cite{2016nH} and nGd~\cite{2012DYB} analysis.}
\centering
\begin{ruledtabular}
\begin{tabular}{lcc}
 & nH & nGd \\ \hline
Coincidence time & [1, 400] $\mu$s & [1, 200] $\mu$s \\
Prompt energy  & (1.5, 12) MeV & (0.7, 12) MeV \\
Delayed energy & (1.9, 2.7) MeV   & (6, 12) MeV \\
Coincidence distance & $<$ 500 mm & N/A \\
\end{tabular}
\end{ruledtabular}
\label{IBDSelection}
\end{table}

To illustrate the influence of this novel background, a toy estimate of \T~is made.
In Ref.~\cite{2016nH}, two results were reported.
The ratio of the observed to the expected number of IBDs at the far hall $R = 0.950 \pm 0.005$.
The value of \T~is $0.071\pm 0.011$.
Taking this background into consideration and setting the background rate to 0.15 per day, $R$ reduces to 0.947.
Assuming very roughly that $1-R$ is proportional to \T, the central value of \T~will increase from 0.071 to 0.075.
Note that this is a very simple estimation just for illustrating that \T~will increase when taking into consideration this novel background.
The proper quantitive assessment of the radiogenic background’s impact on \T~shall be done by the Daya Bay collaboration with all the detailed knowledge and know-how about the experiment.

The simulation has also been performed using detector geometries of RENO and Double Chooz.
The two experiments chose Hamamatsu R7801 low radioactivity PMTs.
The \U~and \Th~contaminations of R7801 PMTs are 50\% to 70\% of R5912 PMTs in Daya Bay.
The oil shields of the two experiments were both designed to be 70~cm, larger than the 50~cm shield of Daya Bay.
Both effects lead to a negligible radiogenic neutron background contribution from PMT glass, i.e., less than 5$\times 10^{-4}$ per day, in RENO and Double Chooz.

\subsection{Discussion}

The above simulation is performed only with a Daya Bay like detector.
The selection cuts are applied only on variables calculated in the stage of detector simulation.
Neither the energy resolution nor the vertex resolution is included.
A detailed simulation study shall be carried out by the experiment with its official simulation software, which has been tuned with data.

It should be noted that only radiogenic neutrons generated in the PMT glass are simulated in the present study.
In many cases, materials rich in fluorine, such as PTFE and Viton, are also frequently used in the detector for supporting and sealing and are closer to the LS target.
The \U~and \Th~contaminations in PTFE are at the ppb level, but those in Viton are several tens ppb.
Radiogenic neutrons generated in PTFE and Viton might not be negligible.
Since these materials' masses and radiopurity levels are not found in the literature for the three experiments, we do not include these neutrons in this study.
We recommend reactor neutrino experiments to examine how many materials rich in fluorine are used in the detector.
Any increase of the radiogenic neutron background, if not corrected for, would lead to a smaller value of \T.

In addition, the secular equilibrium is assumed for \U~and \Th~decay chains in this study.
However, the equilibrium is rarely achieved in most surface and near-surface geological environments.
For example, radon gas emanation during glass production breaks the equilibrium.
On the other hand, the $^{210}$Pb isotope accumulates easily on surfaces of materials.
The disequilibrium effect would introduce systematic uncertainties to the estimate of the radiogenic neutron background.

\section{Summary}
\label{sec5}
In summary, we have found that radiogenic neutrons in peripheral materials could mimic neutrino signals in reactor neutrino experiments and affect the measurement of the neutrino mixing angle $\theta_{13}$.
Setting the \U~and \Th~decay chains to secular equilibrium, the neutron yields and energy spectra of ($\alpha$,~n$_{i}$) are calculated for two commonly used materials, borosilicate glass and PTFE.
Results of the calculation are used in a Geant4 based simulation.
In a neutrino detector like Daya Bay, radiogenic neutrons generated in PMT glass could introduce about 0.15 background events per day in the neutrino dataset detected via neutron capture on hydrogen.
Taking this background rate into consideration, the \T~value measured with this dataset will increase compared to that in Ref.~\cite{2016nH}.
In RENO and Double Chooz, due to the lower radioactive contamination in PMT glass and the thicker oil shielding compared to Daya Bay, radiogenic neutrons from PMT glass have negligible contributions.
In the neutrino datasets detected via neutron capture on gadolinium of the three experiments, the radiogenic neutron background can be safely ignored.

The measurement of the neutrino mixing angle $\theta_{13}$ is of great importance for neutrino physics.
The precision will be dominated by reactor neutrino experiments for decades to come.
We recommend that Daya Bay, RENO, Double Chooz, and JUNO, which will measure the other three neutrino oscillation parameters to sub-percent precision, to carefully examine the radiogenic neutron background in their detectors, including masses and radiopurity levels of detector materials, in particular those rich in boron and fluorine.

\section{Acknowledgement}
This work was supported in part by National Natural Science Foundation of China under Grant No.~11975244, by the National Key R\&D Program of China under Grant No.~2018YFA0404100.
The authors would like to thank Jen-Chieh Peng, Ming-chung Chu, and Bed\v{r}ich Roskovec for carefully reading the manuscript and valuable comments.

\bibliographystyle{apsrev}
\bibliography{RadioNeu}

\end{document}